# Multi-Input Multi-Channel Analyzer (MIMCA) using universal FPGA board

H.Andrianiaina[1*], H.Rongen[2], Raoelina Andriambololona[3], G.Rambolamanana[4], J.B. Ratongasoandrazana[5]

[1] Maintenance and Instrumentation Department, Institut National des Sciences et Techniques Nucléaires (INSTN-Madagascar), B.P.4279, Antananarivo, Madagascar
*hery_andrianiaina5@yahoo.fr*

[2] Zentralinstitut für ElektronikInstitut (ZEL), Forschungszentrum Jülich, 52425 Jülich Germany
*H.Rongen@fz-juelich.de*

[3] Theoretical Physics Department, INSTN-Madagascar,
*raoelinasp@yahoo.fr*

[4] Institut Observatoire et Géophysique d'Antananarivo (IOGA), Antananarivo, Madagascar
*g_rambolamanana@yahoo.fr*

[5]Maintenance and Instrumentation Department, INSTN-Madagascar
*ratongasoarjb@gmail.com*

*Corresponding author

**ABSTRACT**: Instrumentation for nuclear applications is developing very fast, due to fast changing of technology in electronics in connection to Moore's Prediction ("doubling of transistor density integration on an IC for every two years"). The maintenance concept has changed accordingly and moved from repairing at electronic component level to software solving approach, which leads to customize application to fit the local needs. Therefore, Madagascar-INSTN has developed some R&D projects in instrumentation to support and fit to the local needs: one example is the FPGA-based Multi-Input Multi-Channel Analyzer using the UNIO52 standard board from Jülich, which could be used to strengthen the capability and speed-up the routine for radionuclide measurement and analysis of samples. The test results of the MIMCA performance (Linearity, Count-rate accuracy tests) show that the configurable digital system can be used as an alternative issue compared to branded acquisition equipments, which are very expensive and limited technical support from supplier.

## I. Introduction

According to the By-law n° 3961/93, all foodstuffs must be controlled by the Madagascar-INSTN, and must have a non-radioactive certificate, prior to the selling. In that case, the level of radioactivity, especially for artificial radionuclide, must be below the legally accepted values. The INSTN has set up an agreement with the Trade Ministry, from January 1997 regarding the implementation of the regulation.
Mainly the rate of radioactive contaminating element is very low; therefore the sample analysis needs a long time for measurement. As a matter of business, customers need to have their certificate to be delivered as rapid as possible, to allow fast clearance of their goods.
Using a multiple set of equipments would need more investment and would cost a lot in case dedicated and branded instruments (Germanium detector with thick lead shield coupled with MCA).
This paper shows the use of a low-cost universal acquisition board to build a Multi-Channel Analyzer with multi-inputs to monitor in parallel up to four samples

## II. FPGA-Based Multi-Channel Analyzer

### A. *Multi-Channel Analyzer (MCA)*

There are various methods of measuring amplitude of pulses depending of the application, but single channel analysis (SCA) and multi-channel analysis (MCA) methods are mostly used.
Single Channel Analyzer (SCA) is used for counting the number of incoming radiation at selected energy range **[1]**, with two adjustable levels. Only pulses with amplitude falling in between the two levels are counted. Input pulses are observed by two discriminators set at LL (Lower Level) and UL (Upper Level) respectively. MCA method has to be used to have the distribution of pulse heights.

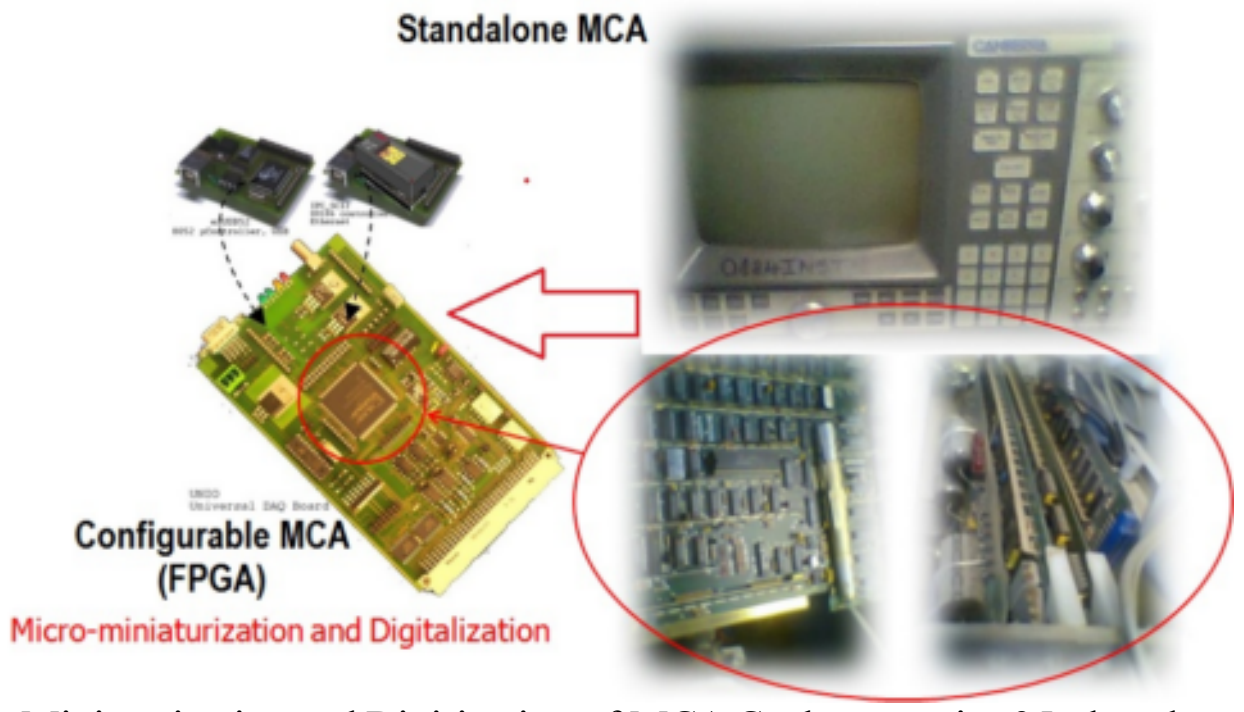

**Fig. 1:** Miniaturization and Digitization of MCA Canberra series 35 plus electronics

Due to the rapid advancement of large integrated circuits, FPGAs (Field Programmable Gate Array) are utilized in many applications to control or to process signals and are used in nuclear instruments. FPGAs are semiconductor devices containing programmable logic and programmable interconnects

### B. *FPGA-based Pulse Height Analysis (PHA) system*

For MCA application, the Pulse Height Analysis (PHA) algorithm has to be described in VHDL and the configuration file has to load into the FPGA. The principle and algorithm of Pulse Height Analysis (PHA) is based on sampling the current input signal from the ADC until it reaches values greater than the programmed Lower Level (sMcaLL), and continuing sampling data until the signals goes lower than Lowest Lower Level reference value (sMcaLLL) or the number of sampled points is more than sMaxPulsLen **[2]**. During the pulse sampling, the FPGA always compares the current value (sMcaPeak) against the last value and holds in a register the maximum of the values (sMcaPeakMax). So after the signal goes lower than Lowest Lower Level this register holds the Pulse maximum. The registered pulse maximum is now compared with Upper Level (sMcaUL). In the case the pulse maximum is greater than Upper Level or the registered pulse length exceeds the maximum length, the pulse will be discarded; if not, an Interrupt for the microcontroller is generated so that the µC can readout the Event-Maximum.

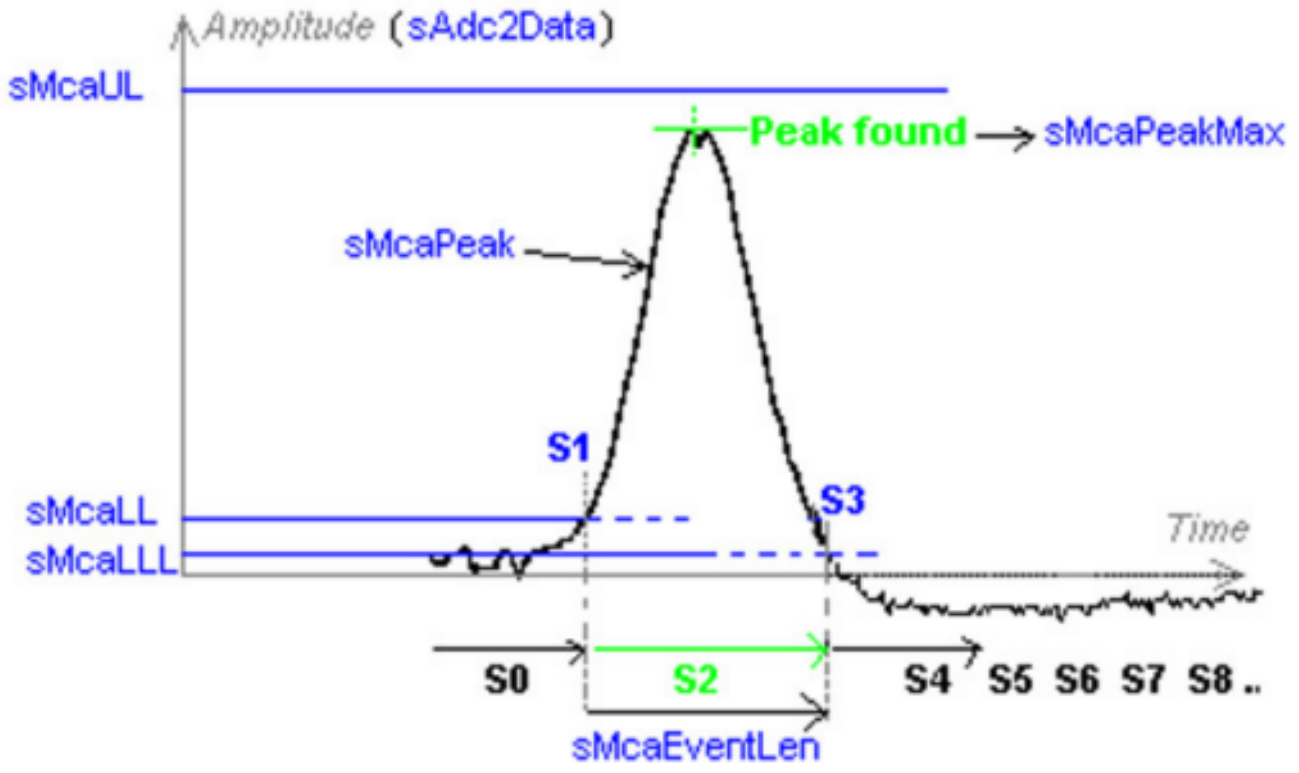

**Fig. 2:** Pulse Height Detection State Machine

## III. The Multi-Input Multi-Channel Analyzer

The system is based on the universal data-acquisition board UNIO52 and LabVIEW graphical programming language for the programming of the application software.

### A. *Configuration of MIMCA onto UNIO52 FPGA board*

The UnIO52 board was developed as a universal data-acquisition and processing board and can be used for nuclear instruments.

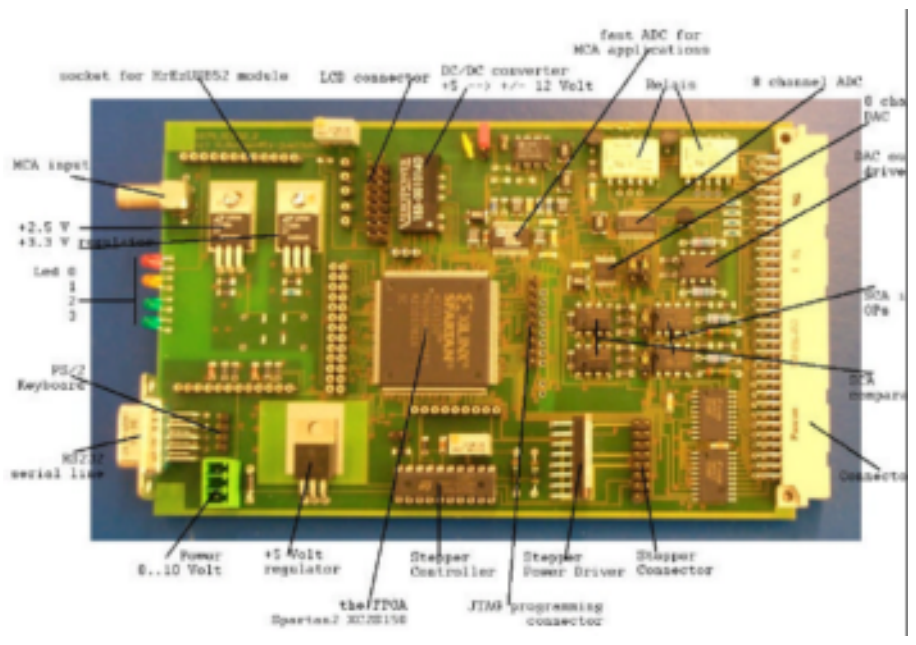

**Fig. 3:** The universal UNIO52 board

Logic blocks have been configured to perform drift control due to voltage or temperature shifting and other complex digital functions such as a peak finder (needed in nuclear instruments) and mathematical functions.
The UnIO52 board consists of a fast ADC, a powerful FPGA and a USB microcontroller to establish the communication with the PC.

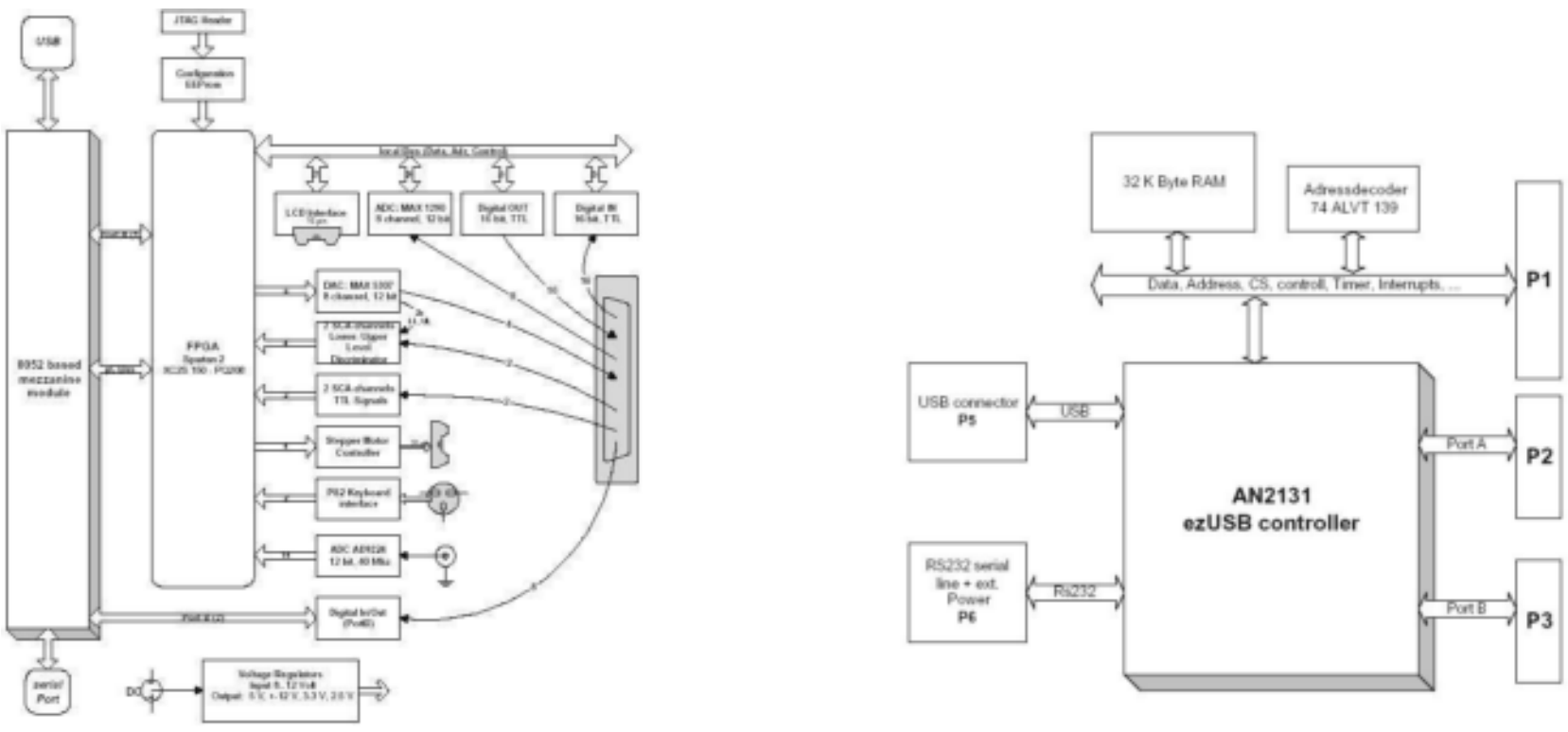

**Fig. 4:** UNIO52 and USB-microcontroller block diagram

All these components are programmable by the user. Therefore this is an ideal platform for the development of a dedicated instrument.

### B. *The MIMCA complete system*

The firmware application code for the ezUSB controller board was developed with Keil C-compiler. The host PC software was designed in the sense that there is a central control for the 4 MCA channel for setting the threshold levels, starting, stopping, clearing the MCA, etc.
Configured UNIO52 boards are assembled in a small 19” Rack, together with 5Volt power supply. The USB cables are fed via a USB hub as one single connection to the PC. The two analog inputs are connected by BNC cables. In the background the nuclear test pulser and a detector system can be seen.

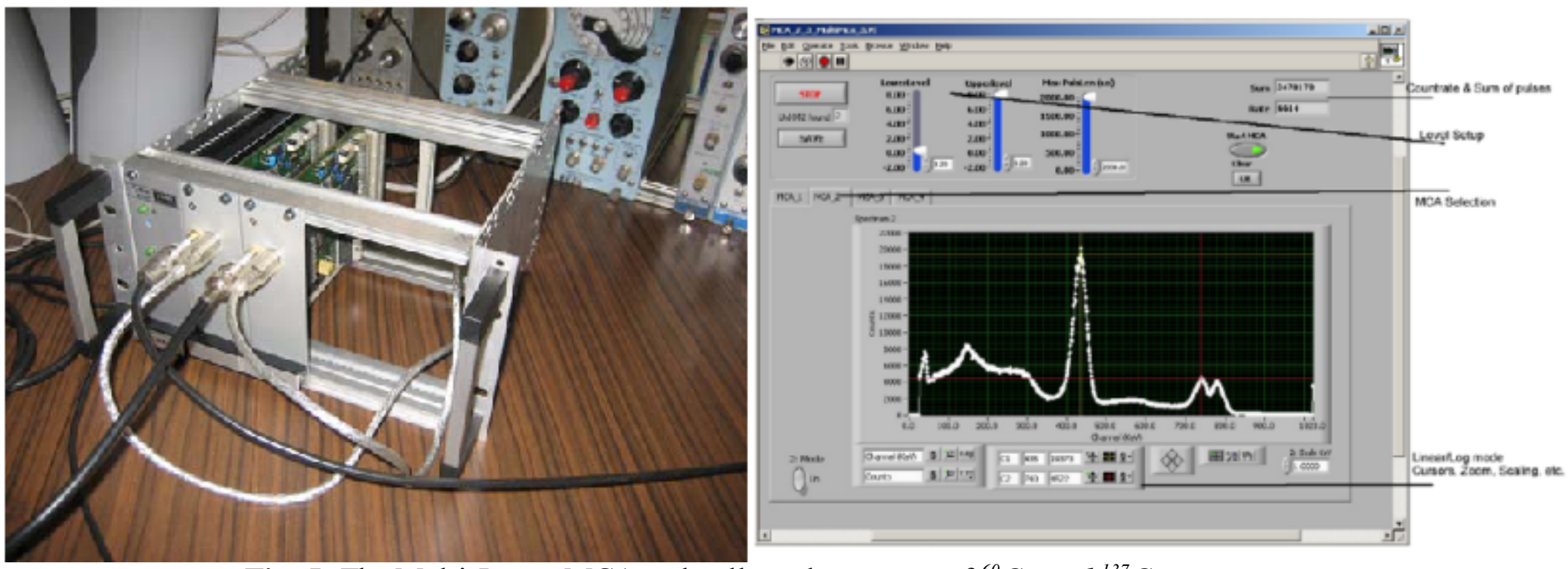

**Fig. 5:** The Multi-Inputs MCA and collected spectrum of $^{60}$*Co and* $^{137}$*Cs point sources*

The 4 spectra’s are shown in 4 graphs which placed in a tab control. By clicking on the “MCA Selection” one of the 4 tabs are selected to show the spectra of the according MCA channel. The MCA’s have a resolution of 1024 channels which were tested with a nuclear pulser. Collected data can be stored to hard disc and can be read by other computing or visualization software such as Matlab and Excel.

## IV. EXPERIMENTAL RESULTS

Digitized spectrum data are stored and incremented periodically inside memory of the UNIO52 board. For further processing (such as peak search), data have to be download into the PC RAM.

### A. *Automatic Peak Search*

An algorithm to latch “Previous integral spectrum” and subtract it from the “new integral spectrum” is implemented with the analyzing software tools. This is to isolate “actual spectrum” from the “integral spectrum”, and will give information about peak channel shifting depending on the amplitude of the analog input signal. This is useful for linearity test and resolution measurement of the system.

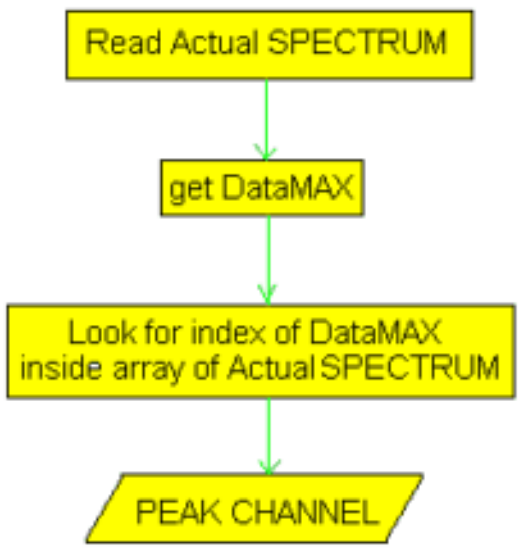

**Figure 6:** Automatic Peak Search algorithm

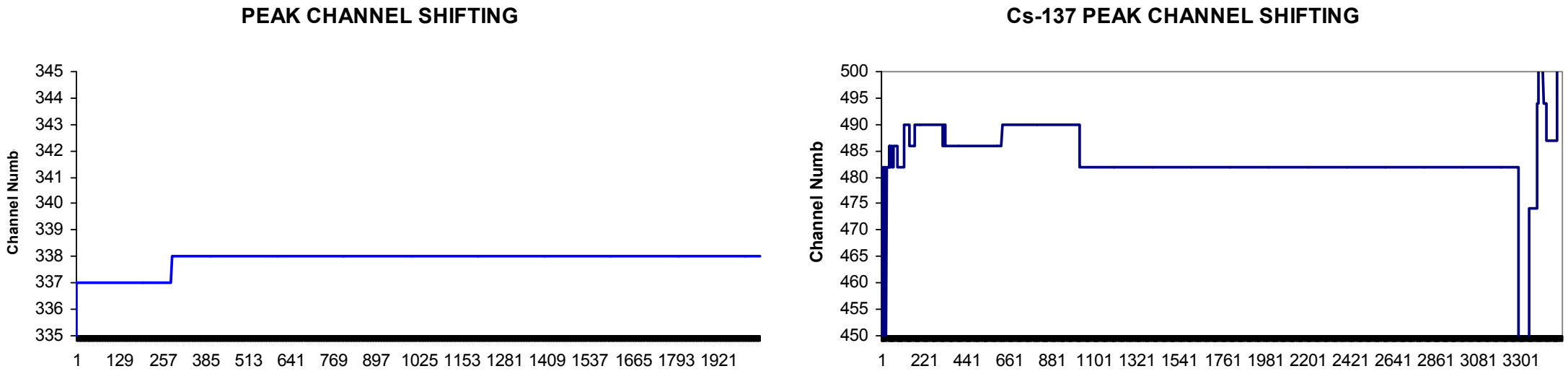

**Fig. 7:** Peak Channel shifting profile using the Automatic Peak Search algorithm

## B. *Linearity and Accuracy*

The main part of a Multi-Channel Analyzer system (MCA), which makes possible the dialogue between the real world of analog quantities (charges released by detector) and digital electronic system is the ADC. The analog-to-digital conversion (ADC) consists of subdividing the input signal range (between $V_{MIN}$ and $V_{MAX}$ ) into $N$ equal parts (ideal ADC), and assigning the input signal amplitude $V_i$ to a channel.

### 1. Linearity

#### *a) Integral Non-Linearity (INL)*

With MCA system, precise energy calibration of a spectrum depends on the linearity of the system. For further data processing in the ideal case the linearity should be as a straight line. Set of pulses with defined amplitudes has been fed to the input of the MIMCA.

The Integral Non-Linearity is given as the maximum channel deviation from the estimated linear regression approximation $CH_{MAX}$ expressed in percent of the total number of channel $N$ **[3]**:

$$INL(\%) = \frac{CH_{Max}}{N} \times 100$$

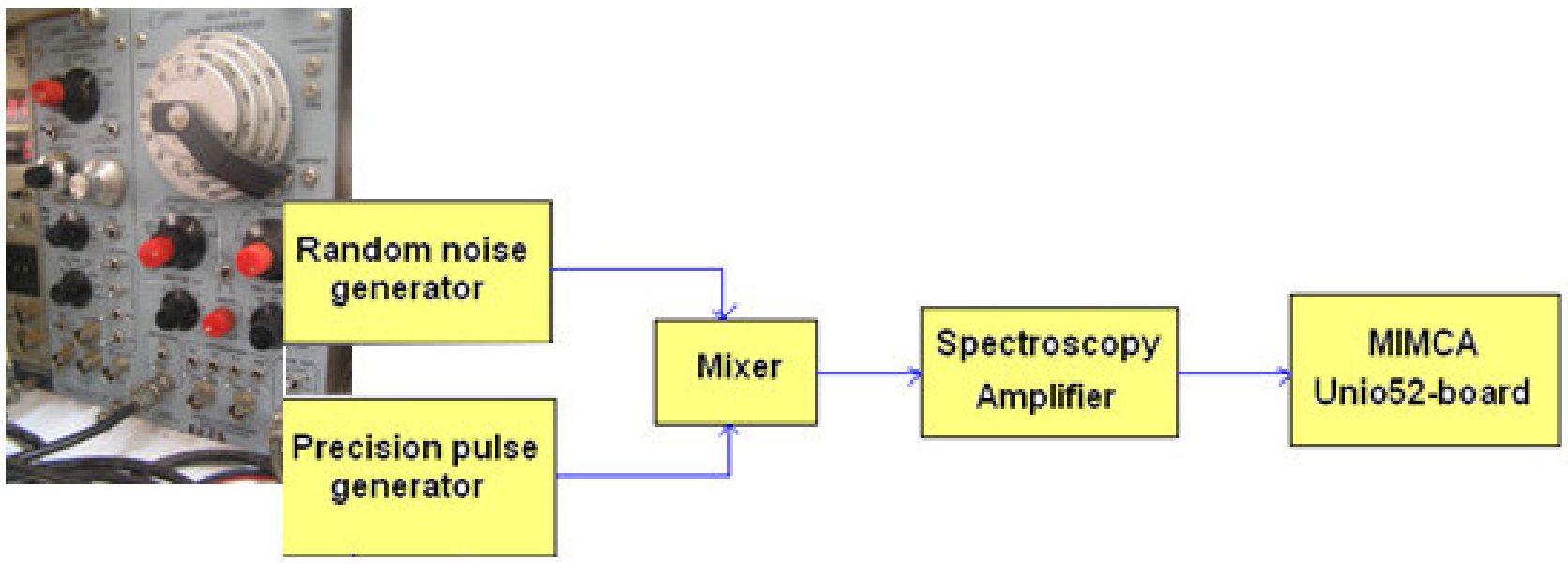

**Fig. 9:** Set-up for INL Test

Table 1 shows collected peak channel location depending on the amplitude (H: height) of the reference signal (peaks in the spectra would have equal distances when the system is linear).

**Table 1:** Peak Height Channel Distribution of MIMCA

| DIAL (*Reference Pulser*) | Signal Height (***H*** in *mV*) | Peak Channel (*CH*) |
|---|---|---|
| 100 | 400 | 46 |
| 200 | 800 | 92 |
| 300 | 1240 | 138 |
| 400 | 1600 | 184 |
| 500 | 2000 | 229 |
| 600 | 2500 | 275 |
| 700 | 2900 | 321 |
| 800 | 3300 | 366 |
| 900 | 3700 | 411 |
| 1000 | 4000 | 457 |
| 1100 | 4500 | 502 |
| 1200 | 5000 | 548 |
| 1300 | 5400 | 593 |
| 1400 | 5800 | 639 |
| 1500 | 6200 | 684 |
| 1600 | 6600 | 729 |
| 1700 | 7000 | 773 |
| 1800 | 7400 | 818 |
| 1900 | 7800 | 863 |
| 2000 | 8200 | 908 |
| 2100 | 8600 | 951 |
| 2200 | 9200 | 998 |
| 2240 | 9400 | 1016 |

For linear regression approximation of the above collected data (channel [CH] function of signal height [H]), the estimated curve equation is the following:

$$[CH] = 0.1089 \times [H] + 7.4389$$

The calculated integral non-linearity values (10% to 90% of full range) are then **+0.7%** and **-0.6% [2].**

*b) Differential Non-Linearity (DNL)*

For ideal system, each channel should have exactly the same width, and will reproduce the amplitude of the incoming signal in an accurate manner. DNL test has been performed using input signal from "sweeping pulser" (precision pulse modulated by triangular signal).

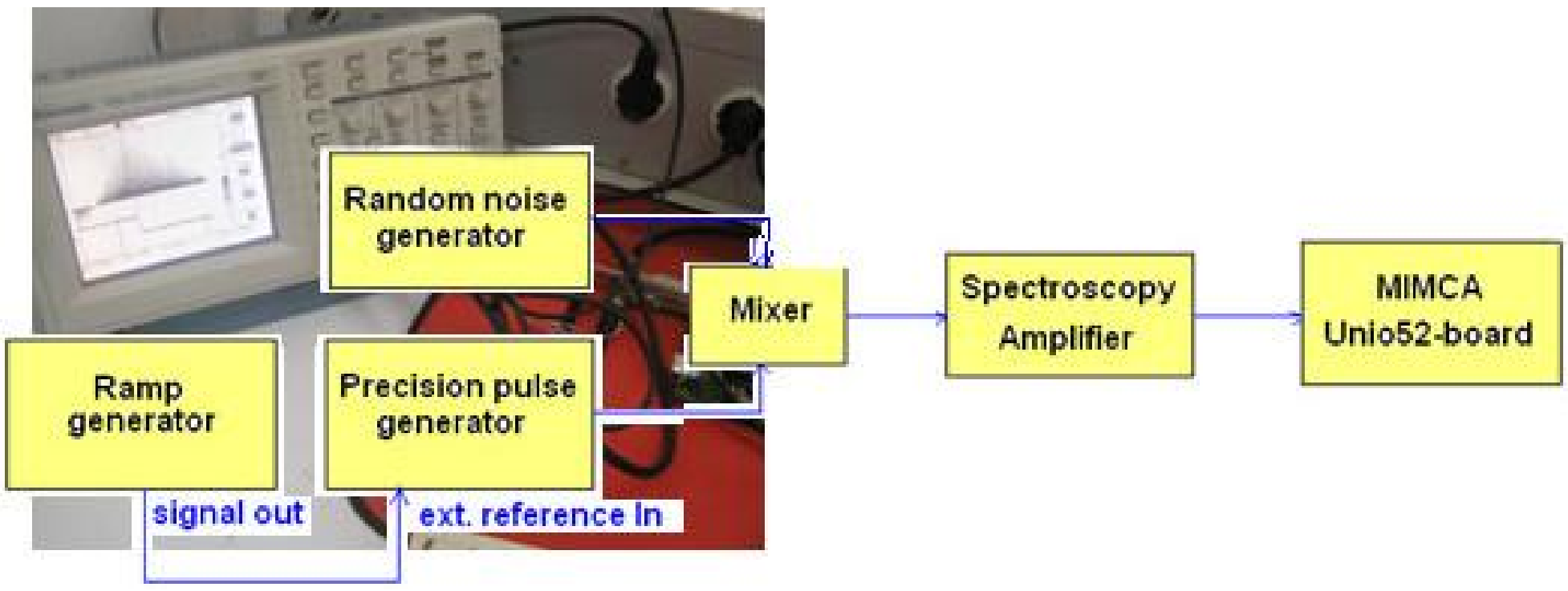

**Fig. 10:** Hardware set-up for DNL test

The settings of the test instruments are as follows:

- Period of ramp generator set to 50sec
- Pulse generator switched to external reference source, repetition rate to 10kHz
- Shaping time of 1µs for the amplifier

Collecting counts over a long period of time (sufficient so that statistical fluctuation can be neglected to the measurement of channel profile), would result uniform and flat distribution of counts in all channels for linear system. Deviation from that uniformity, measures the DNL (expressed in percent) of the MIMCA:

$$DNL(\%) = \frac{\sigma}{\overline{C}} \times 100$$

$\overline{C}$ : Mean value of counts

$\sigma$ : Standard deviation

An overnight measurement (according to the IEC standard 659), gives a mean value of **54160 counts** for each channel. The standard deviation from the mean is **1236 counts**, and the calculated DNL is then **+/- 2.3%**.

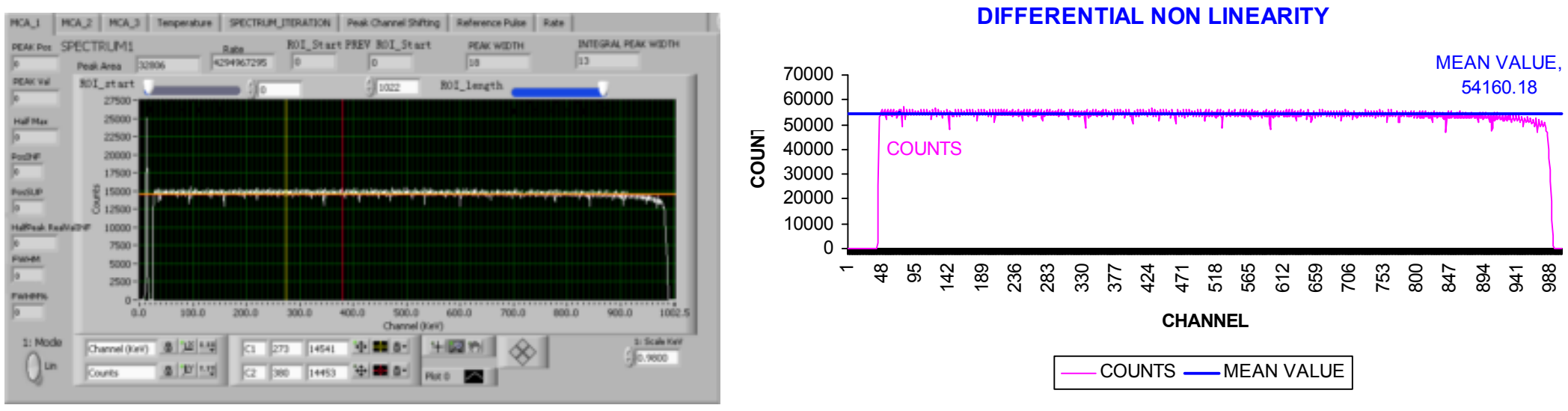

**Fig. 11:** Displayed DNL test result

## 2. Count-rate and Peak Resolution Performance

*a) Count-rate performance:*

For an ideal MCA, all pulses with the same amplitude are stored in one channel, and independent of the count rate.

For testing of the count rate performance, the output of a Random Pulse Generator is connected to the spectroscopy amplifier (CANBERRA model 2020), which fed the MIMCA system.

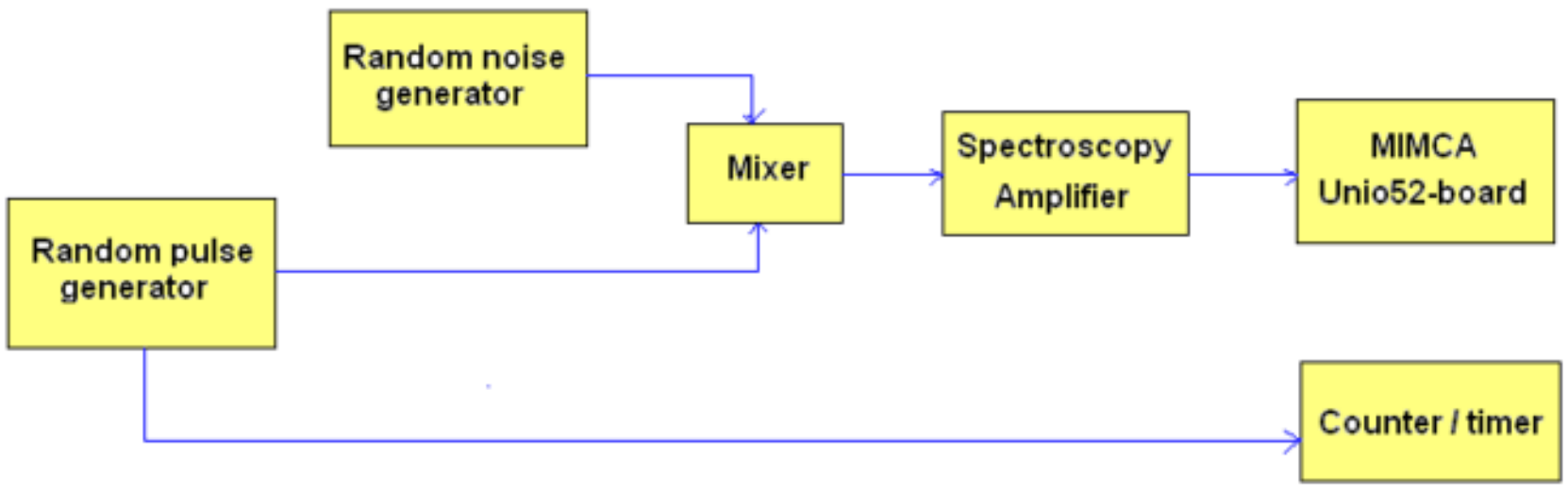

**Fig. 12:** Test set-up for count-rate performance

The Trigger output of the Random Pulse Generator is also connected to a Counter/Timer (CANBERRA model 1776) to serve as reference.
Peak channel of the generated pulse resolved by the MIMCA and function of its count rate is registered, and presented in the following table

**Table 2:** Count-rate Performance of MIMCA

| | Shaping Time | | | |
|---|---|---|---|---|
| | 1us | | 4us | |
| Count-rate [Kcps] | Peak Channel [ch] | Peak shift [%] | Peak Channel [ch] | Peak shift [%] |
| 1 | 689 | 0.00 | 689 | 0.00 |
| 1.5 | 689 | 0,00 | 689 | 0.00 |
| 2 | 689 | 0.00 | 689 | 0.00 |
| 2.5 | 689 | 0.00 | 689 | 0.00 |
| 5 | 689 | 0.00 | 690 | 0.15 |
| 7.5 | 689 | 0.00 | 690 | 0.15 |
| 10 | 690 | 0.15 | 690 | 0.15 |
| 15 | 690 | 0.15 | 690 | 0.15 |
| 20 | 690 | 0.15 | 691 | 0.29 |
| 25 | 691 | 0.29 | 691 | 0.29 |
| 30 | 691 | 0.29 | 691 | 0.29 |
| 39 | 692 | 0.44 | **Double-Peak not resolved** | |
| 50 | **Double-Peak not resolved** | | | |

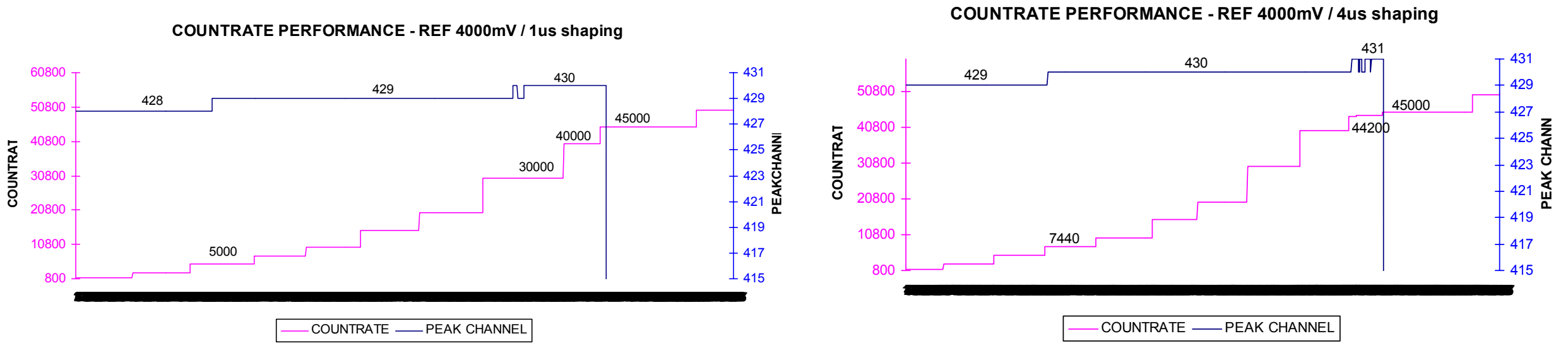

**Fig. 13:** Resolved Peak channel vs. count-rate and shaping of input signal

*b) Peak resolution*

The FWHM resolution of the $^{137}$Cs peak, from collected spectrum within the MIMCA was 21.38%. For comparison, typical FWHM for a MCA35+ CANBERRA system is around 7%.

One of the reason of such difference may be due to the memory limitation made available within the UNIO52 board, which allow only 1K MCA channel resolution

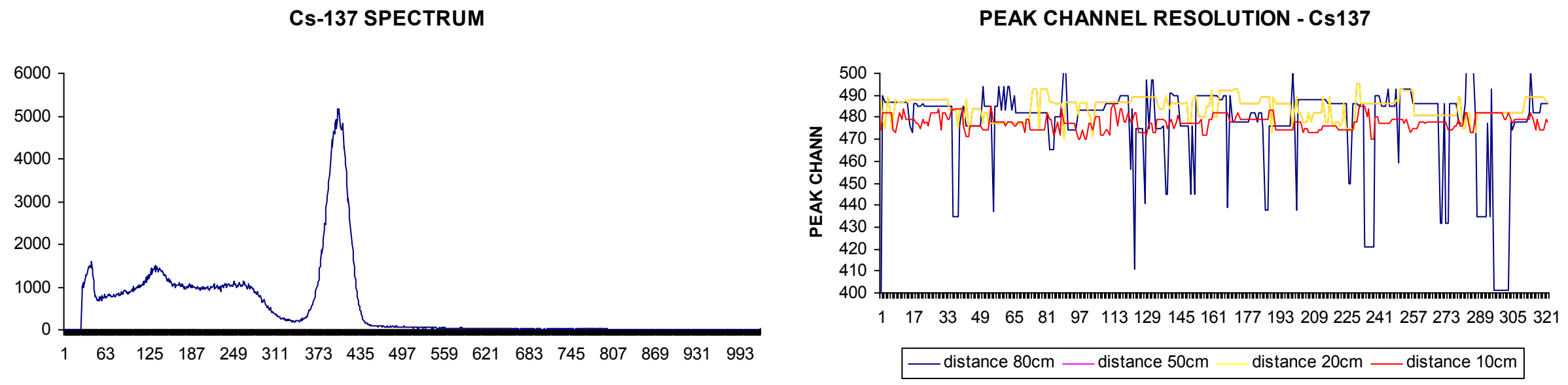

**Fig. 14:** Resolution of $^{137}$Cs Peak-Channel

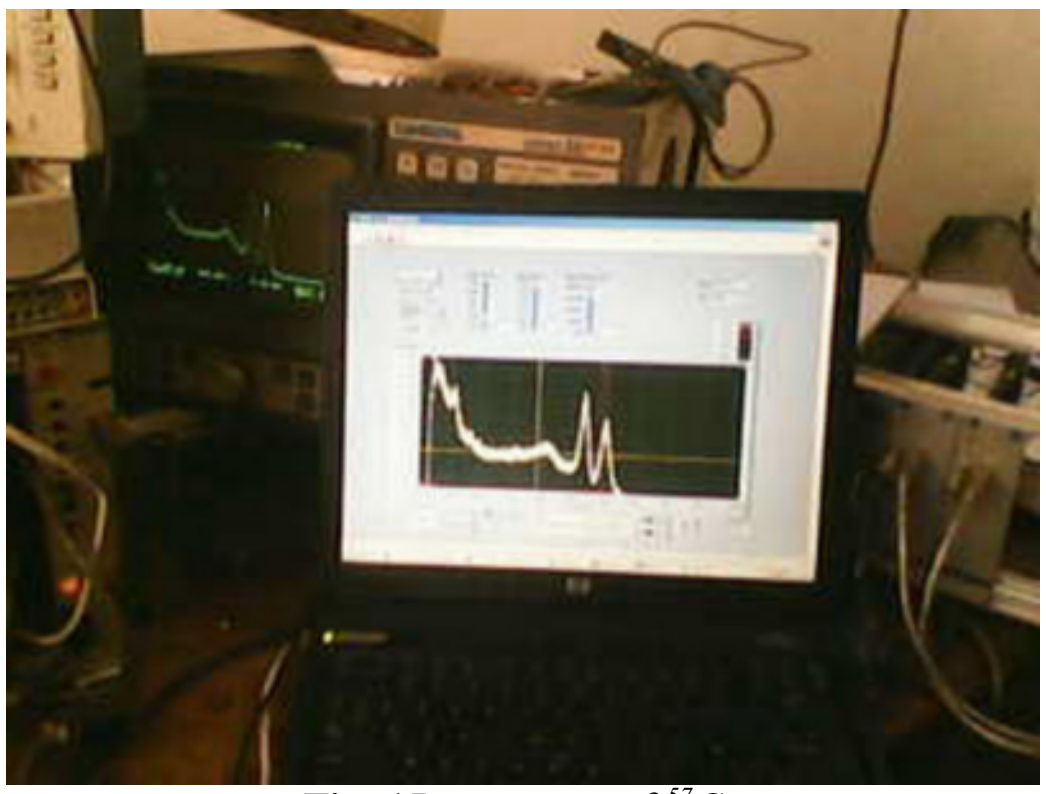

**Fig. 15:** spectrum of $^{57}$*Co*

## V. CONCLUSIONS

- The performance test results show that the built MIMCA (Multi-input MCA) system can used to replace and refurbish dedicated, but expensive Multi-Channel Analyzer.
- The peak resolution within the MIMCA system (21.38% for $^{137}$Cs peak) seems to be worse if compared of those of CANBERRA's (typical FWHM for a MCA35+ CANBERRA system is around 7 %). But the system performance is very sufficient for qualitative analysis of sample.
- Some blocks are still needed to be configured into the UNIO52 board to get full functions of a MCA (Pile-up rejection and dead time correction). Other research projects are scheduled to be done in order to use the development system in nuclear instrumentation improvement (FWHM with low cost spectrometry system).

## VI. ACKNOWLEDGEMENT

Authors wish to extend their great appreciation to:

- The INSTN-Madagascar for setting up the program on refurbishment of scientific equipments, and which gives results to the present work,
- The International Atomic Energy Agency (IAEA-Vienna) for the technical assistance offered through the project MAG/4/004, and supporting with equipments, kits, fellowships and expert missions.

## AUTHOR'S PROFILE

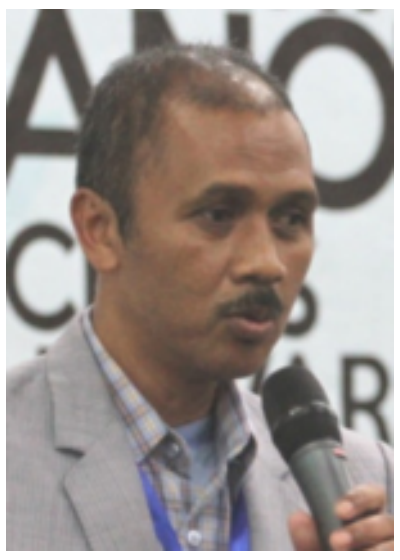

**Hery Andrianiaina:**received the B.S. and M.S. degrees in Nuclear and Applied Physics from the University of Toliara in 1986 and the University of Antananarivo with collaboration of INSTN-Madagascar in 1999, respectively. Since 1994 up to now, he is the Head of the Maintenance and Instrumentation Department at INSTN-Madagascar, and involved on Research and Development activities in nuclear instrumentation.

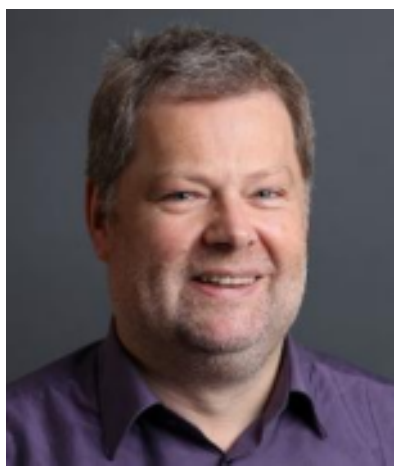

**Dr. Heinz Röngen:** belongs to the "Systeme der Elektronik (ZEA-2)" research group at the Forschungszentrum Jülich. He is working in many projects on Sensor and signal processing systems, and Earth's observation systems (such as the GLObal Limb Radiance Imager of the Atmosphere or GLORIA project, the KOMPSat-3 which is a lightweight Earth observation satellite). He is an owner of some patents on electronic device for measuring biomedical data, and author of many scientific papers.

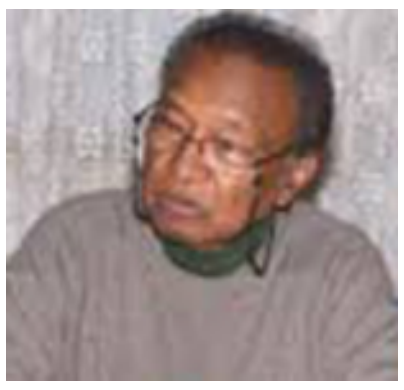

**Prof. Raoelina Andriambololona:** Full Professor, Founder and Director General of INSTN-Madagascar. He is fellow of many learned scientific societies and academics, The World Academy of Sciences for Developing Countries (TWAS), African Academy of Sciences (AAS), New York Academy, Malagasy Academy, American Physical Society, …. He is the author of many, about 250, scientific papers and of 2 university books in Mathematics and Quantum Mechanics. He has served as expert of IAEA, UNESCO, and OAU.

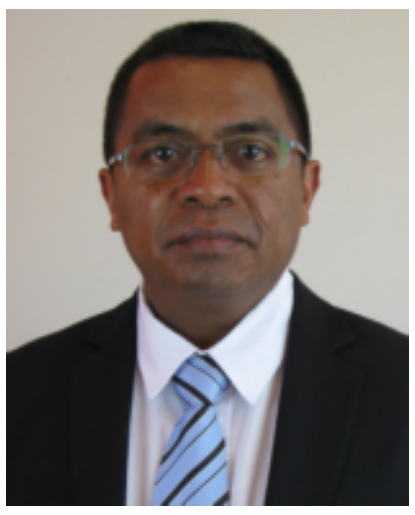

**Prof. Gérard Rambolamanana:** Full Professor of Geophysics, Seismology, Acoustics and Electronics at the University of Antananarivo, Madagascar. He is the Head of the Laboratory of Seismology and Infrasound at the Institute Observatory of Geophysics of Antananarivo (IOGA), at the University of Antananarivo, and has been the Director of the IOGA since 2009. He has been involved in the Square Kilometer Array (SKA) in South Africa since 2009, and worked with the CTBTO since 2011. From 1992 to 2008 he was a Regular Associate then a Senior Associate at the International Centre for Theoretical Physics in Trieste, Italy. He is the author and co-author of several scientific papers related to seismology and infrasound in Madagascar.

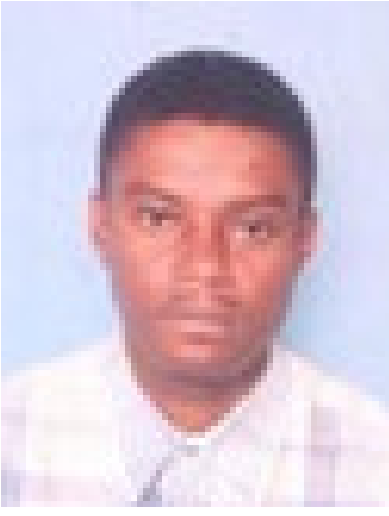

**Jean-Baptiste Ratongasoandrazana:** received his M.S. degrees in Nuclear and Applied Physics from the University of Antananarivo with collaboration of INSTN-Madagascar in 1997. He is working on developing of microcontroller based system at the Maintenance and Instrumentation Department of INSTN-Madagascar since 1998.